\documentclass[12pt]{article}
\usepackage{latexsym,graphicx}
\usepackage{subfigure}
\usepackage{amssymb}
\usepackage{amsmath}
\usepackage{amscd}
\usepackage{amsthm}
\usepackage{float}
\usepackage[left=2cm,top=2.5cm,right=2.5cm,bottom=1.5cm]{geometry}
\usepackage{times}
\theoremstyle{definition}
\newtheorem{definition}{Definition}    
\newtheorem{theorem}{Theorem}[section]       
\usepackage{xcolor}
\begin{document}
\begin{center}
\large{\bf{ A new topological perspective of expanding space-times with applications to cosmology }} \\
\vspace{10mm}
\normalsize{Nasr Ahmed$^{1,2}$, Anirudh Pradhan$^3$ and F. Salama $^{1,4}$}\\
\vspace{5mm}
\small{\footnotesize $^1$ Mathematics Department, Faculty of Science, Taibah University, Saudi Arabia.} \\
\small{\footnotesize $^2$ Astronomy Department, National Research Institute of Astronomy and Geophysics, Helwan, Cairo, 
Egypt}\\
\small{\footnotesize $^3$ Department of Mathematics, Institute of Applied Sciences \& Humanities,
GLA University, Mathura-281 406, Uttar Pradesh, India}\\
\small{\footnotesize $^{1,4}$ Mathematics Department, Faculty of Science, Tanta University, Tanta, Egypt}
\vspace{2mm}

$^{1,2}$E-mail: nasr.ahmed@nriag.sci.eg \\
$^3$E-mail: pradhan.anirudh@gmail.com \\
$^4$E-mail: fatma.salama@science.tanta.edu.eg \\
\end{center}   
\begin{abstract}

We discuss the possible role of the Tietze extension theorem in providing a rigorous topological base to the expanding space-time in cosmology. 
A simple toy model has been introduced to show the analogy between the topological extension from a circle $S$ to the whole space $M$ and 
the cosmic expansion from a non-zero volume to the whole space-time in non-singular cosmological models. A topological analogy to the cosmic 
scale factor function has been suggested, the paper refers to the possible applications of the topological extension in mathematical physics. 

\end{abstract}
{\it Keywords}: Topological extension; cosmic expansion, singularity. \\
{\it Mathematical Subject Classification 2010}: 57Q10, 54C15, 83F05. 

\section{Introduction and Motivation} \label{secintro}

Space-time singularities and the accelerating cosmic expansion are two major challenging problems in modern theoretical physics. 
In order to fully understand and describe what happens near the singularity, a complete quantum theory of gravity is required which 
is still missing. Due to the absence of such a unified theory, the best available option is to investigate toy models in which quantum 
effects are considered \cite{ref1}.  Examples of some candidate theories of quantum gravity are Wheeler-DeWitt theory \cite{ref2}, 
super-string theory \cite{ref3}, loop quantum gravity \cite{ref4}, brane theories \cite{ref5}, and higher-order gravity \cite{ref6}. It has 
 been found in \cite{ref7,ref8} that loop quantum gravity can be very helpful in eliminating singularities. It has also been shown in \cite{ref8d} that quantum gravity effects can remove singularity. Another attempt to overcome the initial cosmological singularity is represented by the oscillatory cosmological models \cite{ref9,ref10,ref11}. It has also 
been shown that cosmological models free from singularity can be obtained in string theories \cite{ref12} and quadratic gravity \cite{ref13}. 
Another singularity-Free Cosmological solution has been obtained in \cite{ref14}. All such theories and attempts show the need for a deeper 
understanding of the space-time singularities from both mathematical and physical sides. In a previous publication \cite{ref15}, Ahmed and 
Rafat explored the problem of space-time singularities from a global topological viewpoint through the topological retraction theory. 
They have investigated the role that can be played by the retraction theory and suggested a mathematical restriction on the formation 
of such singularities. 
\par

Another big mystery is the reason behind the late-time accelerating cosmic expansion \cite{ref16,ref17,ref18} which is still unknown. The existence 
of `Dark Energy' with negative pressure has been suggested as a possible explanation \cite{ref19}. This exotic energy can act as a repulsive 
gravity pushing the universe to expand faster and faster. Modeling the current accelerating expanding universe has been a major subject in
both mathematical and physical cosmology. In the present work, we aim to find a solid topological ground for this observational confirmed 
cosmic expansion. The current failing to understand the transition from decelerating to accelerating cosmic expansion could be due to 
the incomplete understanding of the mathematical nature of the expanding spaces. In the literature, there is a sharp lack of the global 
topological study of expanding spaces against the local geometrical modeling. Some applications of topology in general relativity has 
been studied in \cite{ref20,ref21}. The topological QFT was introduced in \cite{ref22}. To describe deformations in topology we use the 
homotopy \cite{ref23} which has been shown to be very useful in different areas of mathematical physics \cite{ref24,ref25,ref26,ref27,ref28}. 
\par

The space-time global topology is a major problem in mathematical physics and the question of whether this topology is static or variable 
together with the expanding space-time is still open \cite{ref29}. In case it is variable, then the relation between this variable 
space-time topology and the cosmic space-time expansion is unknown. The possibility of a changing space topology was first introduced 
in \cite{ref30}, and the topology change in canonical quantum cosmology has been studied in \cite{ref31}. the quantum changes of 
topology has been discussed through path integral in \cite{ref32,ref33,ref34}. Some approaches to relate topology to cosmology have been 
also studied in \cite{ref35,ref36,ref37,ref38,ref39} (all of them are metric approaches with a local study to the topology change). 
In \cite{ref29}, the expansion of a topological space has been defined using `fractal topology' which is a new suggested topological 
tool through which the detection of continuous deformations of space is allowed. 
\par

In the current work, our aim is to provide a solid topological base to the space-time expansion through utilizing the extension theorem 
in topology. While all previous studies related to the theory of retracts have been pure mathematical studies, Ahmed and Rafat have started 
a series of publications \cite{ref40,ref41,ref42,ref43} in which some applications of this topological theory in mathematical physics have 
been suggested for the first time.  In \cite{ref15}, it has been suggested that the retraction theory in algebraic topology can provide a 
topological description of the gravitational collapse process which leads to the formation of space-time singularities. We have also clarified 
the similarity between a collapsing physical system (a contracting volume in space-time), and the topological retraction. The retraction and 
folding of the higher-dimensional Schwarzchid metric have been investigated in \cite{ref40}. In \cite{ref41}, we have shown the existence of a 
strong connection between the topological retraction and the holographic principle in quantum gravity where the geometry of the hologram 
boundary can be explored through this connection. From this point of view, the retraction theory represents a solid topological base to 
the holographic principle. In \cite{ref42}, we have used the retraction theory \cite{ref43,ref44} to provide rigorous proof to the 
existence of deformations and dimensional reduction in black holes/wormholes, and to explain the topological origin of such 
deformations/dimensional reduction. 
\par

While in the retraction theory space can retract or deformation retract into a subspace, in the current work we need the inverse 
process with the opposite topological concept given by the Tietze extension theorem. Since such topological extension theorem has not 
been applied before to physics, this gives another advantage to the current work where it opens the door for interesting applications 
of this theorem in different areas of mathematical physics. While in the retraction process space $X$ gets continuously shrinked to 
a subspace $A\subset X$, here we are exploiting the inverse process where we can extend a subspace $A$ to a space $X$. Clearly, this 
way of thinking implies that as space-time expands, the number of space-time dimensions increases which are quite acceptable from 
a physical point of view if the big bang theory is correct ( It is generally believed that a dimensional reduction occurs near the 
Planck length \cite{ref45}). 
\par

While the foundations of the theory of retracts were laid by Borsuk \cite{ref46} who introduced the basic notions, his work had its precedents. 
The Tietze extension theorem \cite{ref47} is the most significant among these precedents which proves the extension of continuous functions on a 
closed subset of normal topological space to the entire space. The Tietze Extension Theorem provides general conditions under which it can 
be concluded that extensions exist \cite{ref48}. Specifically, if $A$ is a closed subset of a normal space $X$, and $J\subset R$ is either 
a closed bounded interval, an open interval, or all of $R$, then this Theorem confirms that every continuous $f: A\rightarrow J$ extends to 
a continuous function $F: X\rightarrow J$. Tietze theorem was first proved for metric spaces and then Uryson proved his well-known lemma 
(section 2). For a detailed discussion with an interesting historical review see \cite{ref49}.
\par
The basic motivation behind this work is to find a topological base for the cosmological expansion confirmed by recent observations. 
Let’s see how the current discussion of the expanding space-time is related to previous works. Firstly, the current work discusses the 
global topological side while previous studies are interested in the local geometrical research. There is only one topological study 
on expanding spaces which have been done in \cite{ref29} through introducing the new concept of 'fractal topology', but in the current 
paper we use basic algebraic topological notions and theorems with no need to introduce new concepts. On the other hand, the need for 
a global topological view in describing the expanding space-time has become necessary after the discovery of the challenging problem 
of the accelerating cosmic expansion. In other words, the incomplete understanding of such accelerating expansion motivates the search 
for different mathematical models with wider views. Secondly, based on homotopy aspects illustrated in \cite{ref15}, the current work 
describes an expansion that starts from a non-zero volume and not from a space-time singularity. So, it provides a topological base to a 
singularity-free cosmic expansion where the absence of the initial singularity is a major advantage in cosmology. Thirdly, the paper 
represents the first application of the topological extension theorem in cosmology and mathematical physics.
\par

In the current work, we introduce a toy model for the extension from a circle in a metric space ( and not from a singularity as we 
clarified in \cite{ref15} based on homotopy aspects) to the higher dimensional space.  The paper is organized 
as follows: The introduction and motivation are described in Sect. $1$. The basic definitions and theorems are included in Sect. $2$. 
A quick review of the retraction of the whole space into a lower subspace is given in Sect. $3$. A  review of the retraction method developed 
and used in \cite{ref40,ref41,ref42} is given Sect. $4$ and we refer to the original papers for the details. 
In Sect. $5$, we discuss the extension from $S_{i}$ to $M$ and cosmological analogy to the continuous function $f$ in Tietze theorem. 
The last section includes the conclusion.


\section{Definitions and theorems} \label{sec1}

\theoremstyle{definition}
\begin{definition}
``A subspace $A$ of a topological space $X$ is called a retract of $X$, if there exists a continuous map $r : X \rightarrow A $ 
such that $X$ is open and $r(a) = a$ (identity map), $\forall a \in A$. Because the continuous map $r$ is an identity map from 
$X$ into $A \subset X$, it preserves the position of all points in $A$'' \cite{ref23}.
\end{definition}

\theoremstyle{definition}
\begin{definition}{Deformation retract:}\label{def1}
``A subset $A$ of a topological space $X$ is said to be a deformation retract if there exists a
retraction $r: X \rightarrow A$, and a homotopy $f : X \times [0,1] \rightarrow X $ such that \cite{ref50}: 
$f (x, 0) = x$ $\forall x\in X$, $f(x, 1) = r (x)$ $\forall r\in X$, $f(a, t) = a$ $\forall a\in A, t \in [0,1]$.''
\end{definition}


\subsection{Urysohn's lemma and Tietze extension theorem \cite{ref51}} \label{ss}

\begin{itemize}

\item Let $A$ and $B$ be two disjoint closed subsets of a metric space $X$. Then there exists a continuous function $f: X \rightarrow I$ 
such that $f^{-1}(0)=A$ and  $f^{-1}(1)=B$.

\item Let $F$ be a closed subset of a metric space $X$. Then any continuous function $f: X \rightarrow [-1,1]$ extends over the whole $X$.

\item (\textbf{Urysohn's lemma}) Let $A$ and $B$ be two nonempty disjoint closed subsets of a normal space $X$. Then there exists a continuous 
function $f: X \rightarrow I$  such that $f(A) = 0$ and $f(B) = 1$. This lemma is a key ingredient in the proof of the Tietze extension theorem. 

\item (\textbf{Tietze extension theorem})
While Urysohn's lemma proves that on a normal topological space disjoint closed subsets may be separated by continuous functions, Tietze 
extension theorem proves that such continuous functions extend from closed subsets of normal topological space to the whole space. The 
formal statement is given as follows: Let $A$ be a closed subset of a normal space $X$. Let  $f: A \rightarrow [-1,1]$ be a continuous function. 
Then, $f$ has a continuous function $F: X \rightarrow [-1,1]$ such that $F|_A=f$. (\textbf{The statement of the Tietze theorem remains true if 
we replace the segment $[-1,1]$ by a circle $S^1$} \cite{ref51}, this note is essential for the current work as we will see).

\item \textbf{Tietze extension theorem for metric spaces} 
Let $A$ be a closed subset of a metric space $X$; then every continuous $f: A \rightarrow [-1,1]$ extendes to a continuous 
function $F: X \rightarrow [-1,1]$.

\item The Tietze extension theorem is used to prove the general existence theorem about retractions, details of the proof can be 
found in \cite{ref48}.

\end{itemize}

It is also helpful, before leaving this section, to summarize the work we have done in \cite{ref15}. First, we performed a retraction 
to a $5D$ metric $M$ onto $4D$ circles $S_i \subset M$. Such $4D$ circles can still be retracted to a point. However, by defining the 
appropriate homotopy, the existence of a deformation retract on $M$ has been proved which means that the circles can not continue shrinking 
into a point or a 'singularity'. From a physical viewpoint, such shrinking of the space into a point is analogous to a universe collapsing 
into a singularity. Consequently, the deformation retracts we have proved on $M$ stops the formation of the singularity. In the current work, 
and considering the same cosmological space-time, we are interested in extending the space from a lower-dimensional circle $S$ to the whole 
space. From a physical viewpoint, such extension is analogous to a continuously expanding universe from a non-zero volume. 


\section{A quick review to the retraction of the whole space into a lower subspace.} \label{sec3}

In this section, we briefly summarize the basic idea and main result obtained in \cite{ref15} without rewriting the details of calculations 
again in the current work. We have performed a retraction of the following five-dimensional (5D) cosmological space-time $M$ into subspaces 
considering only the flat case supported by recent observations \cite{ref52,ref53,ref54}:  
\begin{equation} \label{mett}
ds^{2}= B^2 dt^2-A^2 \left(\frac{dr^2}{1-\kappa r^2}+r^2 d \Omega^2 \right)-dy^2,
\end{equation}
where $\kappa$ is the 3D curvature index ($k = \pm 1$, $0$). The reasons behind our choice of this Ricci-flat space-time are: 1- it represents an extension to the FRW solutions \cite{basic1}. 2- the special importance of Ricci-flat manifolds in gravity and geometry \cite{zu}. We then obtained the following relations for the coordinates
\begin{eqnarray} \label{xi}
x_o&=&\pm \left(\frac{2n^3(K-y)^2+n^2(5K^2-2Ky-3y^2)+4n(K+\frac{1}{2}y)^2+K(K+2y)}{2t^{2n}(4n^3-4n^2-n+1)}+C_o\right)^{\frac{1}{2}},\\  \nonumber 
x_1&=&\pm \sqrt{Ar^2+C_1}, \  \  \  x_2=\pm \sqrt{A^2r^2\theta^2+C_2}, \  \  \ x_3=\pm \sqrt{A^2r^2 \sin^2\theta \phi^2+C_3},\\  \nonumber
x_4&=&\pm \sqrt{\frac{1}{2}y^2+C_4},
\end{eqnarray}
where $C_{i}$ are the integration constants. Then, After using Euler-Lagrange equations to explore the geodesics, we obtained the following 
set of equations directly from (\ref{xi}) for $\phi=0$ and $\theta=0$ respectively

\begin{eqnarray} \label{coos1}
x_o^{\phi=0}&=&\pm \left(\frac{2n^3(K-y)^2+n^2(5K^2-2Ky-3y^2)+4n(K+\frac{1}{2}y)^2+K(K+2y)}{2t^{2n}(4n^3-4n^2-n+1)}+C_o\right)^{\frac{1}{2}},
\\  \nonumber
x_1^{\phi=0}&=&\pm \sqrt{Ar^2+C_1}, \  \  \  x_2^{\phi=0}=\pm \sqrt{A^2r^2\theta^2+C_2}, 
\  \  \ x_3^{\phi=0}=\pm \sqrt{C_3}, \  \  \ x_4^{\phi=0}=\pm \sqrt{\frac{1}{2}y^2+C_4}.
\end{eqnarray}
Since
\begin{equation}
ds^2=x_{1}^{2}+x_{2}^{2}+x_{3}^{2}-x_{o}^{2}>0.
\end{equation}
is satisfied, this retraction leads to a circle $S_1 \subset M$ \cite{ref40,ref41,ref42,ref15}. For $\theta=0$, we get
\begin{eqnarray} 
\label{coos2}
x_o^{\theta=0}&=&\pm \left(\frac{2n^3(K-y)^2+n^2(5K^2-2Ky-3y^2)+4n(K+\frac{1}{2}y)^2+K(K+2y)}{2t^{2n}(4n^3-4n^2-n+1)}+C_o\right)^{\frac{1}{2}},
\\  \nonumber
x_1^{\theta=0}&=&\pm \sqrt{Ar^2+C_1}, \  \  \  x_2^{\theta=0}=\pm \sqrt{C_2}, \  \  \ x_3^{\theta=0}=\pm \sqrt{C_3}, 
\  \  \ x_4^{\theta=0}=\pm \sqrt{\frac{1}{2}y^2+C_4}.
\end{eqnarray}
Which also leads to a circle $S_2 \subset M$. So, we have a retraction of $M$ defined as $R : M \Rightarrow S_i ,\,i=1,2$ which 
proves the following theorem
\begin{theorem}
Some types of the geodesic retractions of the 5D cosmological space-time $M$ are circles $S_i \subset M$. 
\end{theorem}


\section{Extension from $S_i$ to $M$ and the cosmological analogy to the continuous function $f$ in Tietze theorem}

Having obtained all the required tools (Theorems, definitions, and proof for the existence of a retraction from $M$ to $S_i$), extending 
the subspace (expanding it) back from $S_i$ to $M$ should now be straightforward and can be considered as a topological base to a cosmological 
expansion from a non-zero volume (the lower-dimensional subspace at the beginning of time) to the whole higher-dimensional space-time (the 
volume of the current universe). 
\par 

We recall that The statement of the Tietze extension theorem remains true if we replace the segment 
$[-1,1]$ by a circle $S$ \cite{ref51} (section \ref{ss}). Starting from a circle $S$ in a metric space $M$, and choosing the spherical 
coordinates as a frame of work along with the time $t$, the following inequality is valid in $M$
\begin{equation}
ds^2=x_{1}^{2}+x_{2}^{2}+x_{3}^{2}-x_{o}^{2}>0.
\end{equation}
From the existence of the retraction (section \ref{sec3} ), there exists a set of constants $C_i$ such that the coordinates 
$x_1$, $x_2$, $x_3$, $x_4$ and $x_o$ can be expressed as in (\ref{coos2}), i.e. :
\begin{eqnarray} 
x_o^{\theta=0}&=&\pm \left(\frac{2n^3(K-y)^2+n^2(5K^2-2Ky-3y^2)+4n(K+\frac{1}{2}y)^2+K(K+2y)}{2t^{2n}(4n^3-4n^2-n+1)}+C_o\right)^{\frac{1}{2}},
\\  \nonumber
x_1^{\theta=0}&=&\pm \sqrt{Ar^2+C_1}, \  \  \  x_2^{\theta=0}=\pm \sqrt{C_2}, \  \  \ x_3^{\theta=0}=\pm \sqrt{C_3}, 
\  \  \ x_4^{\theta=0}=\pm \sqrt{\frac{1}{2}y^2+C_4}.
\end{eqnarray}
Which still can be extended, by appropriate choices for the constants, and generalized to the case when $\theta \neq 0$ to get the form (\ref{xi})
\begin{eqnarray} 
x_o&=&\pm \left(\frac{2n^3(K-y)^2+n^2(5K^2-2Ky-3y^2)+4n(K+\frac{1}{2}y)^2+K(K+2y)}{2t^{2n}(4n^3-4n^2-n+1)}+C_o\right)^{\frac{1}{2}},\\  \nonumber
x_1&=&\pm \sqrt{Ar^2+C_1}, \  \  \  x_2=\pm \sqrt{A^2r^2\theta^2+C_2}, \  \  \ x_3=\pm \sqrt{A^2r^2 \sin^2\theta \phi^2+C_3},\\  \nonumber
x_4&=&\pm \sqrt{\frac{1}{2}y^2+C_4}.
\end{eqnarray}
So, we are back to the $5D$ metric of the expanding space-time (\ref{mett}). An important point is that what makes the appropriate choices 
of constants always guaranteed is the existence of a deformation retract from $M$ to $S_i$ described in section \ref{sec3}. The general existence 
theorem about retraction can always be proved using the Tietze extension theorem as we have mentioned in section \ref{sec1}  \cite{ref48}.
\par

\begin{figure}[H]
  \centering            
  \subfigure[]{\label{F555}\includegraphics[width=0.36\textwidth]{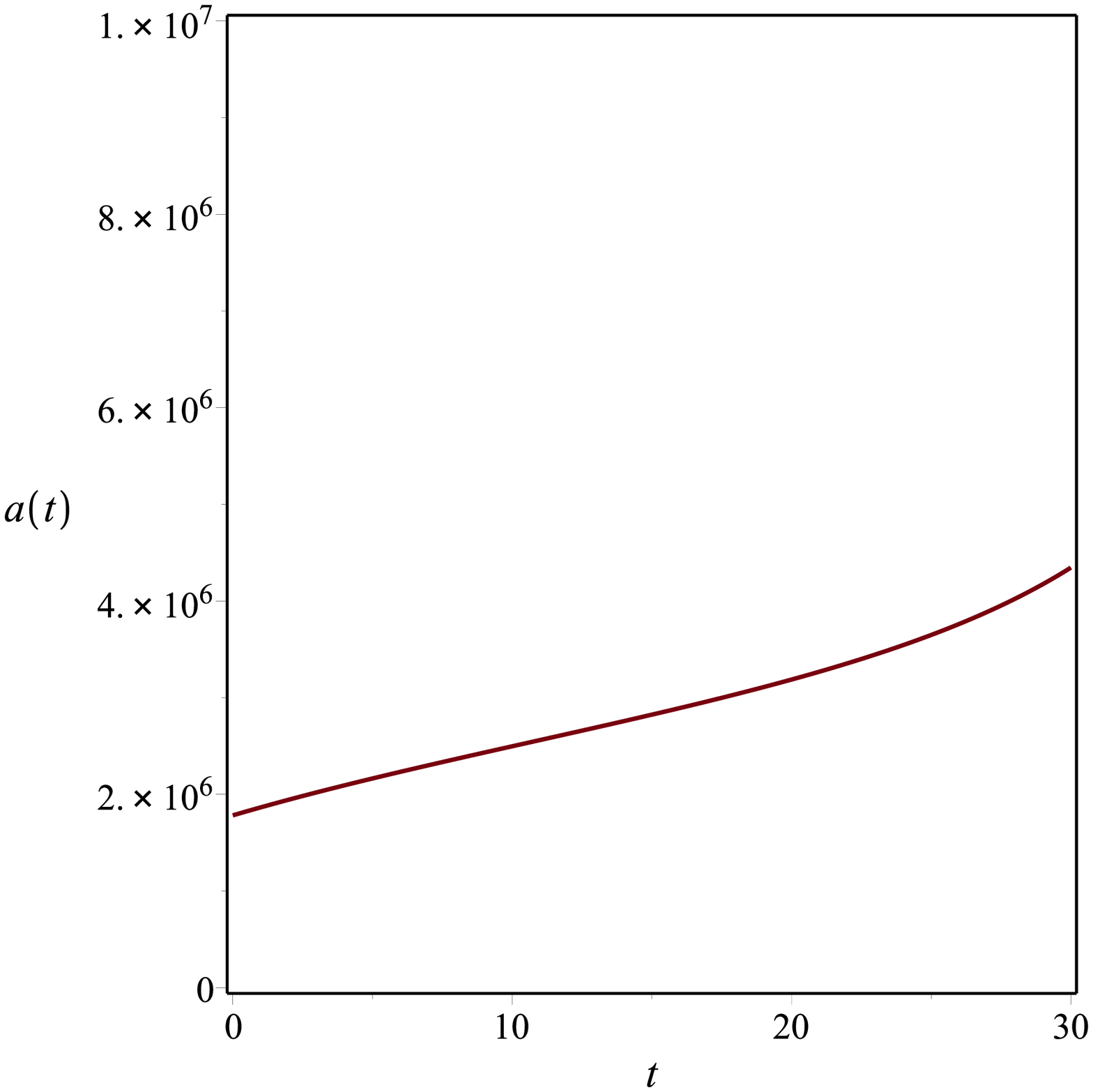}}
  \subfigure[]{\label{F63}\includegraphics[width=0.36\textwidth]{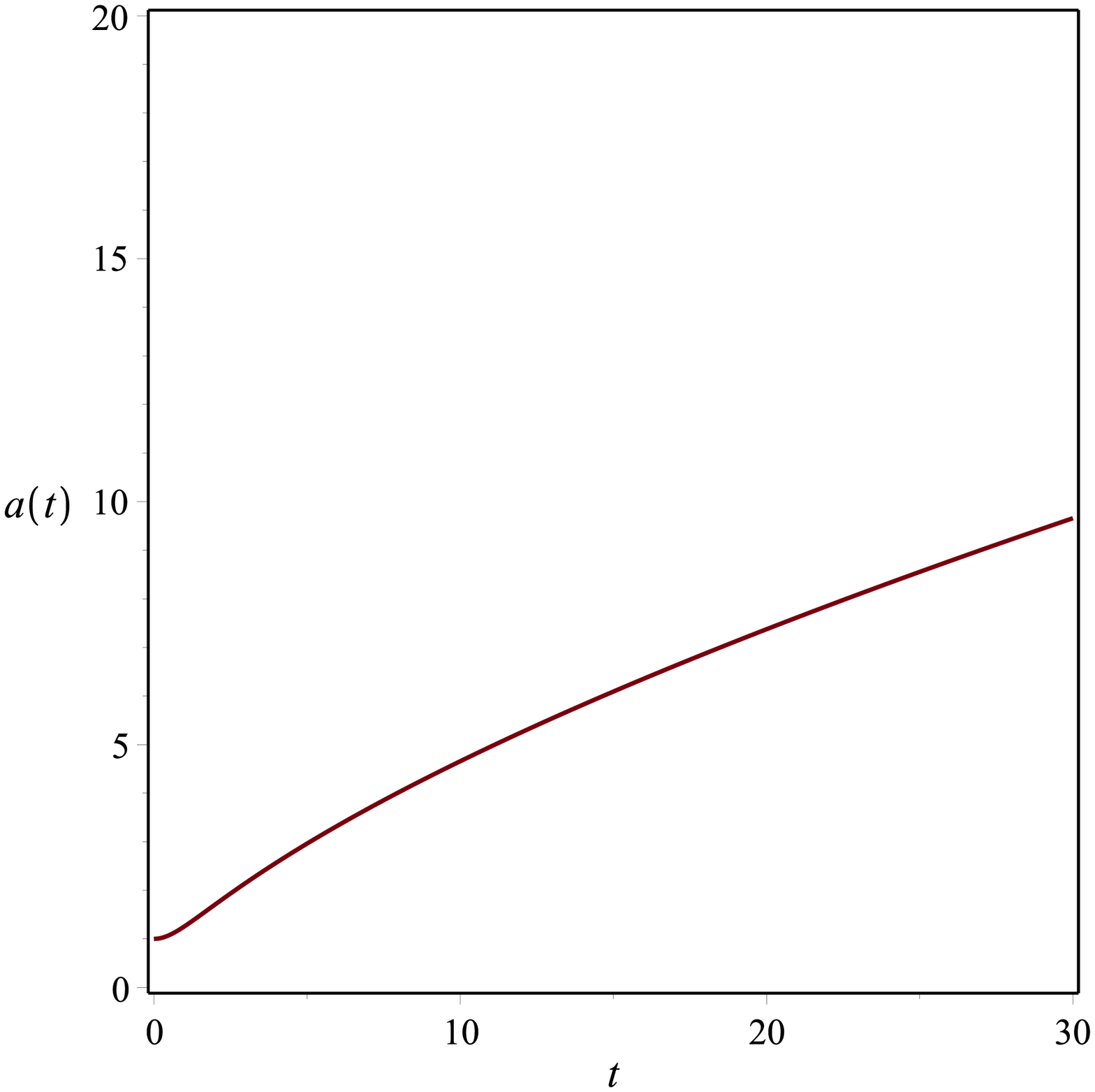}} \\
  \label{fig:cassimir55}
\end{figure}

Recalling that the metric (\ref{mett}) represents an extension to FLRW solutions, this topological extension from simple circles to the 
whole cosmological space-time provides a rigorous topological base for any non-singular FLRW cosmological model in which the universe 
starts the expansion from a non-zero volume. A non-singular cyclic cosmological model has been introduced in \cite{ref11} where the 
universe expands from a non-zero volume with a scale factor given by   
\begin{equation} 
\label{a}
a(t)=A \exp \left[ \frac{2}{\sqrt{c^2-m^2}}\arctan \left(\frac{c\tan \left(\frac{kt}{2}\right)+m}{\sqrt{c^2-m^2}}\right)\right]
\end{equation}
where $A$ and $c$ are integration constants. The scale factor (\ref{a}), as a function of cosmic time, corresponds topologically to the 
continuous extendable function $f$ in Tietze extension theorem. This cosmic function grows from a non-zero volume, where its initial radius 
can be easily calculated from the model to the whole universe before shrinking back again to the initial radius. The evolution of $a(t)$ is 
shown in figure \ref{F555} and we can see that it has a non-zero value at the beginning of time $t=0$. The cosmological model in \cite{ref55} 
is another FLRW type model free from the initial singularity developed in the framework of General Relativity with a perfect fluid. The 
singularity-free solution of the model is represented by the scale factor function as
\begin{equation} \label{aa}
a(\tau)= a_o\left[1+\left(\frac{\tau}{T_o}\right)^2\right]^{\frac{1}{3(1-\omega)}}
\end{equation}
Where $a_o$ and $T_o$ are constants, and $\omega$ is the equation of state parameter. The evolution of $a(\tau)$ is shown in figure \ref{F63} 

Cosmological models free from initial singularity may provide solutions to the problems of standard cosmology such as the flatness problem, 
the homogeneity problem, and the generation of primordial perturbations (see for example \cite{ref56}). The initial singularity is a major 
undesirable feature in the standard FLRW cosmology where no geometrical or physical description of space-time is available at the beginning 
of time. For this reason, there are many modified theories of gravity in the literature that tried to eliminate the initial Big-Bang 
singularity problem. Examples of such generalized gravity theories are: higher-order gravity \cite{ref6}, loop quantum gravity \cite{ref58}, 
superstring \cite{ref59} the brane-world gravity \cite{ref60}, and others. Some theories succeeded to introduce cosmological scenarios 
in which the cosmic evolution can be extended through the initial singularity such as the cyclic scenario \cite{ref61} and the pre-big-bang 
scenario \cite{ref62}. In \cite{ref63}, a simple modification of General Relativity theory has been considered where there are no space-time 
singularities at the classical level in FRW and Kasner universes. This opens the door to having a gravitational theory free from singularities 
in general. \par
Although the analysis introduced in this work has been done based on the Ricci-flat metric (\ref{mett}) which represents 
an extension to the FRW solutions, This analogy is general and can be still valid for other non-singular cosmological models. For example, 
a more realistic singularity-free theory for early and late-time accelerated expansion which covers whole universe history has been introduced 
in \cite{real1}.


\section{Conclusion}

Following previous studies where some applications of the deformation retract have been suggested, in the current work, we have 
studied the analogy between the topological extension and cosmic expansion. A toy model has been introduced using a cosmological 
metric where the expansion starts from a non-zero volume. The paper represents the first application of the Tietze extension theorem in 
mathematical physics and opens the door for more applications in other physical systems.\par

Our main point of view behind developing such analogy is to look for a deeper understanding to the mathematical nature of expanding space-times. It is known that implementing new mathematical definitions and theorems into physics increases our ability to see the problem from different sides which motivates exploring new ideas. Because topology covers the global view with rigorous definitions, there have been so many successful attempts to utilize topological notions in Modern physics (we have mentioned some examples in the introduction). The need for a global topological description of expanding space-times has become important after the discovery of the challenging problem of the accelerating cosmic expansion. The incomplete understanding of such accelerating expansion could be due to an incomplete understanding to the mathematical nature of expanding spaces (In the literature, there is a sharp lack of the global topological study of expanding spaces against the local geometrical modeling). 

Although we have clarified the connection between the topological extension and the study of expanding spaces by introducing a simple toy model, this opens the door for more applications of this connection. Suggesting such a new connection can help to understand the expanding space-times from a different mathematical side and, consequently, attacking some related open questions from a different view. many pure topological concepts, such as homotopy, have been used in modern physics and found to be so useful in providing proofs and simplifying calculations. 

One of the main achievements of the current work is that it represents the first application of the topological extension theorem in Physics.  The analysis we have provided is general and not restricted to cosmology, we have suggested a topological base for all expanding space-times which can be used in other fields of physics.   

\section*{Acknowledgment}
We are so grateful to the reviewer for his many valuable suggestions and comments that significantly improved the paper.


\begin{thebibliography}{000}

\bibitem{ref1} 
R. Torres and F. Fayos, Singularity free gravitational collapse in an effective dynamical quantum spacetime, {\it Phys. Lett. B} {\bf 733} (2014) 169-175.
\bibitem{ref2} 
B. S. DeWitt, Quantum theory of gravity, the canonical theory, {\it Phys. Rev.} {\bf 160} (2976) 1113-1148.

\bibitem{ref3} 
J. Polchinski, String duality, {\it Rev. Mod. Phys.} {\bf 68} (1996) 1245.
\bibitem{ref4} 
C. Kiefer, Quantum Gravity, International Series of Monographs on Physics, Third Edition, (2012) 1-432 (Oxford, UK: Clarendon).
\bibitem{ref5} 
L. Randall and R. Sundrum, Large mass hierarchy from a small extra dimension, {\it Phys. Rev. Lett.} {\bf 83} (2999) 3370.

\bibitem{ref6} 
S. Nojiri and  S. D. Odintsov, Unified cosmic history in modified gravity: from $F(R)$ theory to Lorentz non-invariant models, 
{\it Phys. Rep.} {\bf 505} (2011) 59.

\bibitem{ref7} 
W. Struyve, Loop quantum cosmology and singularities, {\it Sci. Rep.} {\bf 7} (2017) 8161.

\bibitem{ref8} 
A. Ashtekar, T. Pawlowski and and P. Singh, Quantum nature of the big bang, {\it Phys. Rev. Lett.} {\bf 96} (2006) 141301.

\bibitem{ref8d} Emilio Elizalde, Shin'ichi Nojiri, Sergei D. Odintsov, Late-time cosmology in (phantom) scalar-tensor theory: dark energy and the cosmic speed-up, Phys.Rev.D 70 (2004) 043539.
\bibitem{ref9} 
J. Khoury, B. A. Ovrut, P. J. Steinhardt and N. Turok, Ekpyrotic universe: Colliding branes and the origin of the hot big bang, 
{\it Phys. Rev. D} {\bf 64} (2001) 123522.


\bibitem{ref10} 
L. A. Boyle, P. J. Steinhardt and N. Turok, New duality relating density perturbations in expanding and contracting Friedmann cosmologies, 
{\it Phys Rev. D} {\bf 70} (2004) 023504.

\bibitem{ref11} 
N. Ahmed and Sultan Z. Alamri, A cyclic universe with varying cosmological constant in $f(R, T)$ gravity, 
{\it Can. J. Phys.} {\bf 97} (2019) 1075-1082.

\bibitem{ref12} 
J. C. Fabris and R. G. Furtado, Singularity-free cosmological solutions in string theories, arXiv:gr-qc/0211096 (2002).

\bibitem{ref13} 
P. Kanti, J. Rizos and K. Tamvakis, Singularity-free cosmological solutions in quadratic gravity, {\it  Phys. Rev. D} {\bf 59} (1999) 083512.
\bibitem{ref14} 
L. Fernandez-Jambrina, Singularity-free cylindrical cosmological model, {\it Class. Quant. Grav.} {\bf 14} (1997) 3407.
\bibitem{ref15} 
N. Ahmed and H. Rafat, Space-time singularities and the theory of retracts, {\it Phys. Scr.} {\bf 94} (2019) 085215. 

\bibitem{ref16} 
S. Perlmutter {\it et al.}, Measurements of omega and lambda from 42 high-redshift supernovae,
{\it Astrophys. J.} {\bf 517} (1999) 565-586.
\bibitem{ref17} 
W. J. Percival {\it et al.}, The 2dF Galaxy Redshift Survey: The power spectrum and the matter content of the universe, {\it Mon. Not. Roy. Astron. Soc.} 
{\bf 327} (2001) 1297.
\bibitem{ref18} 
D. Stern, R. Jimenez, L. Verde, M. Kamionkowski and S. A. Stanford, Cosmic chronometers: Constraining the equation of state of 
dark energy. I: H(z) measurements, {\it J. Cosm. Astrop. Phys.} {\bf 2010} (2010) 008.

\bibitem{ref19} 
N. Ahmed and Sultan Z. Alamri, A stable flat universe with variable cosmological constant in $f(R, T)$ gravity, {\it Res. Astron. Astrophys.} 
{\bf 18} (2018) 123; 
N. Ahmed and Sultan Z. Alamri, Cosmological determination to the values of the pre-factors in the logarithmic corrected entropy-area relation, 
{\it Astrophys. Space Sci.} {\bf 364} (2019) 100. 
\bibitem{ref20} 
A. Chamblin, Some applications of differential topology in general relativity, {\it J. Geom. Phys.} {\bf 13} (1994) 357.
\bibitem{ref21} 
A. Borde, Topology change in classical general relativity, preprint arXiv: gr-qc/9406053 (1994).
\bibitem{ref22} 
M. Atiyah, Topological quantum field theories, {\it Inst. Hautes Etudes Sci. Publ. Math.} {\bf 68} (1988) 175.
\bibitem{ref23} 
L. C. Kinsey, Topology of Surfaces (Springer Verlag, New York, 1993).
\bibitem{ref24}
N. D. Mermin, The homotopy groups of condensed matter physics, {\it J. Math. Phys.} {\bf 19} (1978) 1457.
\bibitem{ref25} 
G. Toulouse, A lecture on the topological theory of defects in ordered media: How the old theory was leading to paradoxes, and 
how their resolution comes within the larger frameworks of homotopy theory. Lecture Notes in Physics 115 (1980). (Springer Berlin Heidelberg).
\bibitem{ref26} 
V. Turaev, Homotopy field theory in dimension and group-algebras, arXiv: math/9910010 (1999).
\bibitem{ref27} 
D. R. Finkelstein, Homotopy approach to quantum gravity, {\it Int. J. Theor. Phys.} {\bf 47} (2008) 534.
\bibitem{ref28} 
M. Benini, A. Schenkel and L. Woike, Homotopy theory of algebraic quantum field theories, {\it Lett. Math. Phys.} {\bf 109} (2019) 1487-1532.
\bibitem{ref29} 
H. Porchon, Topological expansion, study and applications, arXiv:1211.3365[math.GM] (2012).
\bibitem{ref30} 
J. A. Wheeler, On the nature of quantum geometrodynamics, {\it Ann. Phys.} {\bf 2} (1957) 604.
\bibitem{ref31} 
V. A. DeLorenci, N. Pinto-Neto, and I. D. Soares, Topology change in canonical quantum cosmology, {\it Phys. Rev. D} {\bf 56} (1997) 3329–3340.
\bibitem{ref32} 
S. Coleman, Black holes as red herrings: topological fluctuations and the loss of quantum coherence, {\it Nucl. Phys. B} {\bf 307} (1988) 864.
\bibitem{ref33} 
F. Dowker, J. P. Gauntlett, S. B. Giddings, and G. T. Horowitz, Pair creation of extremal black holes and Kaluza-Klein monopoles, 
{\it Phys. Rev. D} {\bf 50} (1994) 2662.
\bibitem{ref34} 
S. Giddings and A. Strominger, Loss of incoherence and determination of coupling constants in quantum gravity, 
{\it Nucl. Phys. B} {\bf 307} (1988) 854-860.
\bibitem{ref35}  
G. F. R. Ellis, Topology and cosmology, {\it Gen. Relativ. Gravit} {\bf 12} (1971) 7-21.
\bibitem{ref36}  
J. R. GottIII, D. H. Weinberg, and A. L. Melott, A quantitative approach to the topology of large scale structure, {\it Astrophys. J.} 
{\bf 319} (1987) 1-8.
\bibitem{ref37} 
J. R. GottIII, M. Dickinson, and A. L. Melott, The sponge-like topology of large scale structure in the Universe, {\it Astrophys. J.} 
{\bf 306} (1986) 341-357.

\bibitem{ref38} 
J. Levin, {\it et al.}, The topology of the universe: the biggest manifold of them all, {\it Class. Quantum Grav.} {\bf 15} (1998) 2689.
\bibitem{ref39}
G. D. Starkman, Topology and cosmology,{\it Class. Quantum Grav.} {\bf 15} (1998) 2529.
\bibitem{ref40} 
N. Ahmed and H. Rafat, Retract and folding of the 5D Schwarzchid field, {\it  Bull. Math. Analy. Applic.} {\bf 7} (2015) 10-19.
\bibitem{ref41} 
N. Ahmed and H. Rafat, Topological origin of holographic principle: application to wormholes, {\it Int. J. Geom. Meth. Mod. Phys.} {\bf 15} (2018) 1850131. 
\bibitem{ref42} 
N. Ahmed and H. Rafat, Existence of deformations and dimensional reduction in wormholes and black holes. 
{\it Int. J. Geom. Methods Mod. Phys.} {\bf 15} (2018) 1850197.
\bibitem{ref43}
Czes Kosniowski, A First Course in Algebraic Topology (Cambridge University Press, 1980); 
E. Bick and F. D. Steffen, Topology and geometry in physics (Springer, 2005).
\bibitem{ref44}
A. Hatcher, Algebraic topology (Cambridge University Press, 2002).

\bibitem{ref45} 
M. Rinaldi, Observational signatures of pre-inflationary and lower dimensional effective gravity, {\it Class. Quantum Grav.} {\bf 29} (2012) 085010.
\bibitem{ref46} 
K. Borsuk, Sur les retracts, {\it Fund. Math.} {\bf 17} (1931) 152-170.
\bibitem{ref47} 
H. Tietze, Uber Funktionen die aufeiner abgeschlossenen Menge stetig sind, {\it J. Reine Angew. Math.} {\bf 145} (1915) 9-14.
\bibitem{ref48} 
C. Adams and R. Franzosa, Introduction to Topology: Pure and Applied. Pearson, 1st edition (2007).
\bibitem{ref49}
I. M. James, History of Topology. North Holland, 1st edition (2006).

\bibitem{ref50} 
W. S. Massey, Algebric topology an introduction (New York, 1967).

\bibitem{ref51} 
O. Viro, O. A. Ivanov, N. Y. Netsvetaev, and  V. M. Kharlamov, Elementary Topology: Problems Textbook (American Mathematical Society, 2008).

\bibitem{basic1} Liu H Yand Wesson P S 2001 Universe Models with a Variable Cosmological "Constant" and a "Big Bounce", Astrophys. J. 562 1.
\bibitem{zu} Cheng L and Zhu 2013 On the Perelman’s reduced entropy and Ricci flat manifolds with maximal volume growth A. Math. Ann. 356 1107.

\bibitem{ref52} 
H. Yand Liu and P. S. Wesson, Universe models with a variable cosmological "constant" and a "Big bounce" {\it Astrophys. J.} {\bf 562} (2001) 1.
\bibitem{ref53} 
P. de Bernardis, P. A. R. Ade, J. J. Bock {\it et al.}, A flat universe from high-resolution maps of the cosmic microwave background 
radiation, {\it Nature} {\bf 404} (2000) 955.
\bibitem{ref54} 
D. N. Spergel, L. Verde, H. V. Peiris, {\it et al.}, First year wilkinson microwave anisotropy probe (WMAP) observations: 
determination of cosmological parameters, {\it Astrophys. J. Supp.} {\bf 148} (2003) 175.

\bibitem{ref55} 
P. Peter and Pinto-Neto Nelson, Cosmology without inflation, {\it Phys. Rev. D} {\bf 78} (2008) 063506.

\bibitem{ref56} 
S. E. P. Bergliaffa, Nonsingular cosmological models, {\it Revista Mexicana de Astronomia  Astrofísica, Conf. Series} {\bf 40} (2011) 5.

\bibitem{ref58} 
M. Bojowald, Loop quantum cosmology, Living Reviews in Rel. {\bf 8} (2005) 11.
\bibitem{ref59} 
J. Polchinski, String duality, {\it Rev. Mod. Phys.} {\bf 68} 1245 (1996).
\bibitem{ref60}
L. Randall and R. Sundrum, Large mass hierarchy from a small extra dimension, {\it Phys. Rev. Lett.} {\bf 83} 3370 (1999); 
L. Randall and R. Sundrum, An alternative to compactification, {\it Phys. Rev. Lett.} {\bf 83} 4690 (1999).
\bibitem{ref61} 
M. Gasperini and G. Veneziano,  The pre-big bang scenario in string cosmology, {\it Phys. Rept.} {\bf 373}(2003) 1-212.
\bibitem{ref62} 
J. Khoury, P. J. Steinhardt, and N. Turok, Inflation versus cyclic predictions for spectral tilt, {\it Phys. Rev. Lett.} {\bf 91} (2003) 161301.
\bibitem{ref63} 
A. H. Chamseddine and V. Mukhanov, Resolving cosmological singularities, {\it JCAP} {\bf 1703} 009 (2017).

\bibitem{real1} E. Elizalde, S. Nojiri, S. D. Odintsov, L. Sebastiani \& S. Zerbini, Nonsingular exponential gravity: A simple theory for early and late-time accelerated expansion, Phys. Rev. D 83 (2011). 

\end{thebibliography}
\end{document}